\input{aipcheck}

\documentclass[
    ,final            
      ]
  {aipproc}

\usepackage{aas_macros}

\layoutstyle{8x11double}
 
\renewcommand\XFMtitleblock{%
  \XFMtitle
  \let\XFMoldpar\par
  \def\par{\XFMoldpar\def\par{\space 
             (on behalf of the VERITAS Collaboration)\XFMoldpar}}%
   \XFMauthors
   \let\par\XFMoldpar
   \XFMaddresses
   \XFMabstract
   \vspace{5pt}%
   \XFMkeywords
   \XFMclassification
 }

\begin{document}

\title{VERITAS Blazar Observations - Recent Results}


\author{Peter Cogan}{
  address={3600 University Ave, Physics Dept, Montreal, QC H3A 2T8, Canada}
}

\begin{abstract}
We present the discovery of very high energy (VHE) gamma-ray emission
from the high-frequency-peaked BL Lac object 1ES 0806+524 (z=0.138)
and the intermediate-frequency-peaked BL Lac object W Comae (z=0.102)
with VERITAS. VHE emission was discovered from these objects during
the 2007/2008 observing campaign, with a strong outburst from W Comae
detected in mid-March, lasting a few days. Quasi-simultaneous spectral
energy distributions are presented, incorporating optical (AAVSO), and
X-ray (Swift/RXTE) observations. We also present the energy spectrum
of the distant BL Lac (z=0.182) 1ES 1218+304 which was detected by
VERITAS during the 2006/2007 observing campaign. The energy spectrum
is discussed in the context of different models of absorption from the
diffuse extragalactic background radiation. We present multiwavelength
observations of the blazar Markarian 421 (z=0.03), including a strong
flare initially detected by the Whipple 10m gamma-ray
telescope. Finally we present a broadband spectral energy distribution
for 1ES 2344+514 (z=0.044) which is successfully fit using a one zone
synchrotron self-Compton model.
\end{abstract}

\maketitle


\section{Introduction}

VERITAS is a ground-based array of imaging Cherenkov telescopes with
sensitivity in the 50 GeV to 50 TeV range. The array if located in
southern Arizona at an altitude of 1280m. There are four 12m diameter
telescopes in the array of the Davies Cotton design. Each telescope
uses a 499 photomultiplier tube camera to image the brief shower of
Cherenkov photons generated by the electromagnetic cascade which
results when a high-energy gamma ray strikes the atmosphere.

Active galaxies are believed to harbour supermassive black holes at
their center surrounded by a hot accretion disk. Highly dynamic
collimated jets of relativistic particles stream away from the plane
of the host galaxy. Particle interactions within the jet produce
electromagnetic radiation across many wavebands, including TeV gamma
rays. Blazars are that subset of active galaxies whose jet is orientated
in the direction of earth. Such objects provide unique laboratories to
study acceleration mechanisms at extreme energies. 

With a wide energy range and excellent point-source sensitivity,
VERITAS is the most sensitive instrument of its kind in the northern
hemisphere and is well suited to observations of blazars. There are
three main areas of interest that are studied using blazar
observations with VERITAS. i) To determine the dominant gamma-ray
production mechanisms in blazars ii) To understand blazar populations
and expand the range of studied objects to low-frequency peaked BL Lac
objects, intermediate-frequency peaked BL Lac objects and flat
spectrum radio quasars iii) To understand the role of the infrared
component of the extragalactic background light in the determination
of intrinsic energy spectra.

In this paper we discuss the discovery of two new objects in the
blazar class, and report on multiwavelength studies of several
established sources.

\section{Discovery of TeV Gamma-Ray Emission from 1ES 0806+514}

\begin{figure}
  \includegraphics[height=.3\textheight]{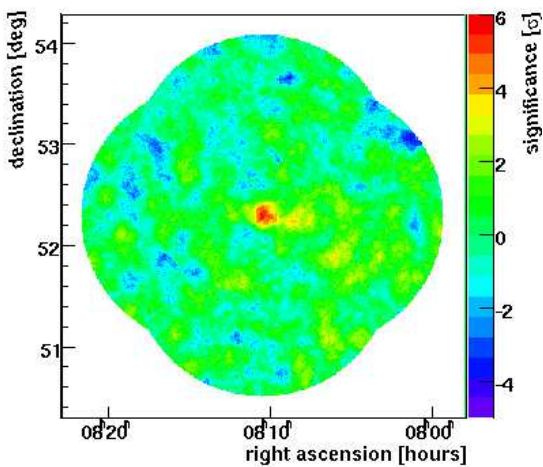} \caption{Two
  dimensional map of statistical significance centered on 1ES
  0806+524. This map indicates that the extent is consistent with a
  point source and there are no other significant TeV sources in the
  vicinity of 1ES 0806+524.}

 \label{fig:1ES0806}
\end{figure}

Deep VERITAS observations of the blazar 1ES 0806+514 (z=0.138) from 2006 to 2008
reveal a weak steady emission above 300 GeV \cite{2008ATel.1415....1S}
(see Figure \ref{fig:1ES0806}). Observations were carried out during
the construction of VERITAS and incorporate data using two, three and
four telescopes. The integral flux above 300 GeV is approximately 1\%
of the Crab Nebula flux. A detailed analysis of these data will be the
subject of a forthcoming VERITAS publication.

\section{Discovery of TeV Gamma-Ray Emission from W Comae}

The intermediate-peaked BL Lac object W Comae (z=0.102) was initially
discovered in the radio and later X-ray energy bands. At gamma-ray
energies, it was detected by EGRET up to 25 GeV with no sign
of a spectral cutoff.

\begin{figure}
  \includegraphics[height=.3\textheight]{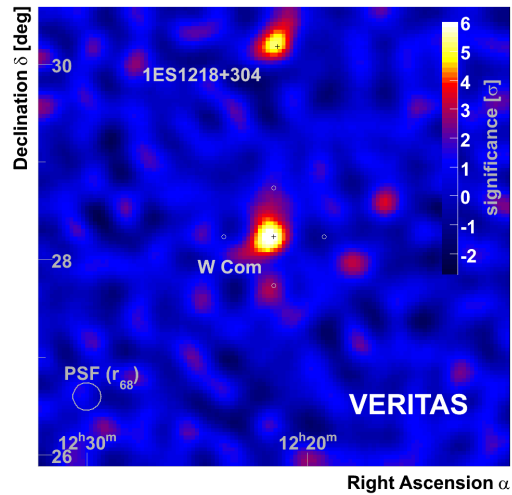} \caption{Two-dimensional
  map of statistical significance of the region centered on W
  Comae. The TeV blazar 1ES 1218+304 is visible to the North of
  W Comae.  }

 \label{fig:WComaeMap}
\end{figure}

VERITAS observed W Comae from January to April 2008 for a total of
39.5 hours. A total statistical excess of $
4.9 \sigma$ (see Figure \ref{fig:WComaeMap}) is recorded during the
observing period, making W Comae the first intermediate-peaked BL Lac
object to be detected in the TeV regime. The majority of the signal
is derived from a flare \cite{2008ATel.1422....1S} over four nights in
March 2008 where the source is detected with a statistical
significance of $6 \sigma$. The differential energy spectrum is well
described by a power law with spectral index $3.81 \pm
0.35_\mathrm{stat} \pm 0.34_\mathrm{sys}$ and flux normalisation
$\left(2.00 \pm 0.31_\mathrm{stat}\right)\times
10^{-11} \mathrm{cm}^{-2} \mathrm{s}^{-1} \mathrm{TeV}^{-1}$. The
total flux above 200 GeV corresponds to roughly 9\% of the Crab Nebula
flux \cite{2008ApJL...648..L73}.

The light curve for the flare nights is fit by the function
$\Phi\left(t\right)=\Phi_0\times \exp\left(-\left(t-t_0\right)^2/\sigma_t^2\right)$
with the flare occurring at $t_0=54538.6 \pm 0.2$ MJD and with the
characteristic time scale $\sigma_t=1.29 \pm 0.28$ days.  No
significant flux variations were measured during the individual
nights, and all other observations during the 39.5 hour exposure yield
no signal. A second strong outburst from W Comae was detected by
VERITAS in June 2008 with an integral flux which exceeded twice the
level of the earlier flare \cite{2008ATel.1565....1S}. Analysis of
these data are ongoing.

Simultaneous SWIFT observations of W Comae were obtained in response
to the detection of flaring behaviour with
VERITAS \cite{2008ATel.1582....1V}. A total of 11.6 hours of SWIFT
UVOT and XRT data are available. The SWIFT data were analysed using
the HEAsoft and XRTPIPELINE tools

\begin{figure}
  \includegraphics[height=.3\textheight]{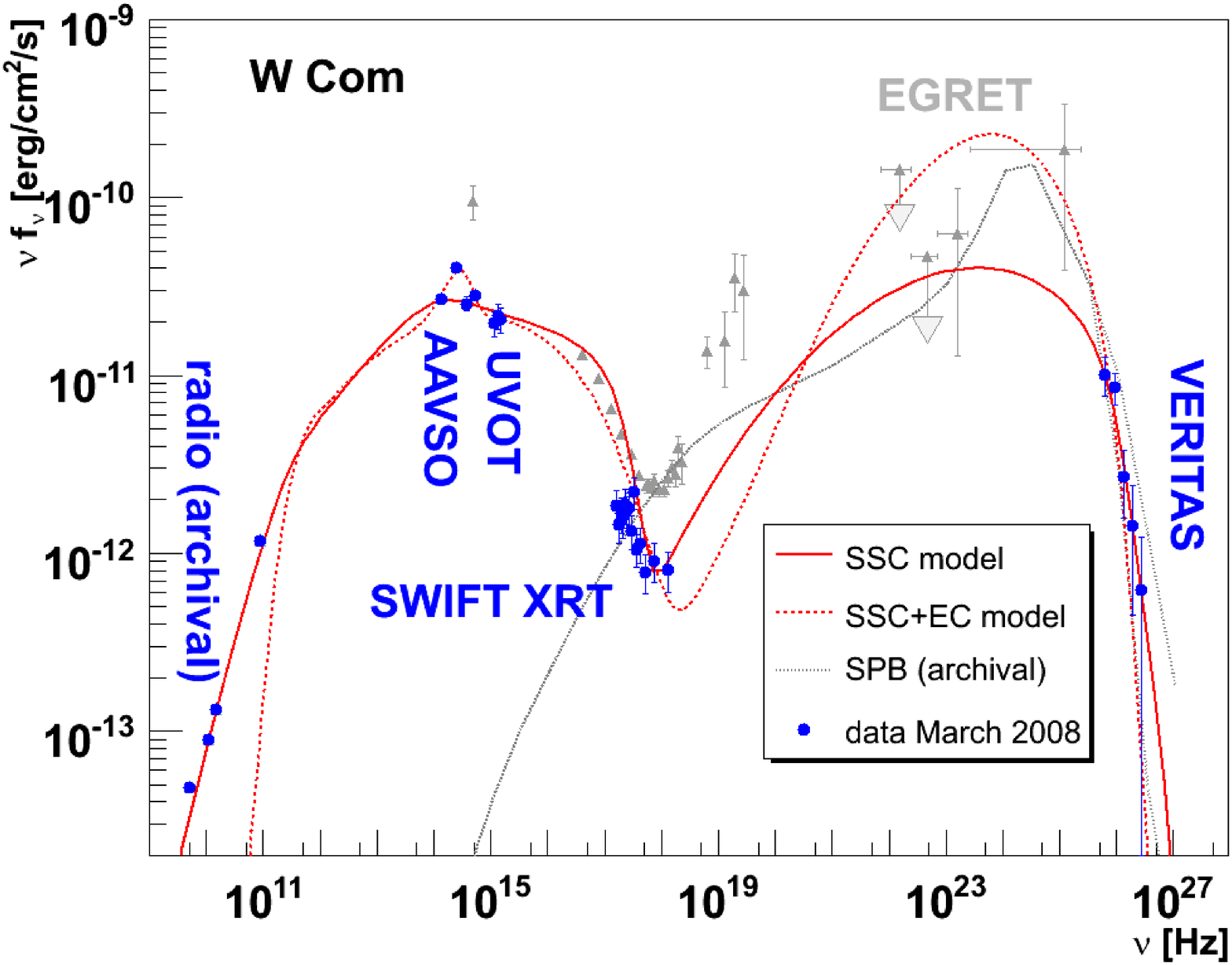} 
  \caption{Spectral
  energy distribution of W Comae comprising archival radio data, AAVSO
  data, SWIFT UVOT and XRT observations, archival EGRET data and
  VERITAS data. Fits to the data are shown for a synchrotron
  self-Compton (with and without an external-Compton component) and an
  archival hadronic synchrotron proton model. }

 \label{fig:WComae}
\end{figure}

The VERITAS and SWIFT data are combined with archival radio data,
optical AAVSO data and archival EGRET data points to produce a
broadband spectral energy distribution (SED) (see
Figure \ref{fig:WComae}). The AAVSO data were obtained in response to the
Astronomer's Alert telegram from VERITAS regarding the gamma-ray
outburst from W Comae.

The SED can be modeled with either a single-zone synchrotron
self-Compton, an external-Compton model or a hadronic synchrotron
proton model. The single-zone synchrotron self-Compton model imposes
the requirement of a relatively weak magnetic field of $\sim 0.007$G
due to the large separation between the first and second peaks in the
SED and the relatively low X-ray flux. The external-Compton model can
also fit the data, but the required magnetic field has a more
realistic value of 0.3 G. See \cite{Beilicke} in these proceedings for
further details.

\section{VERITAS Observations of 1ES 1218+304}

\begin{figure}
  \includegraphics[height=.3\textheight]{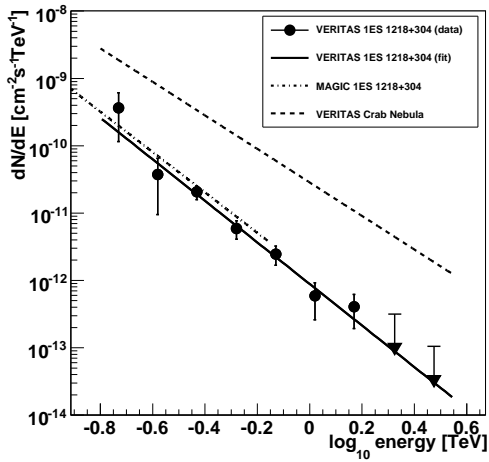} 
  \caption{
  TeV spectrum of 1ES 1218+304 measured with VERITAS - a comparison is
  shown with the spectrum measured by MAGIC, and to the spectrum of
  the Crab Nebula as measured by VERITAS} \label{fig:1218}
\end{figure}

The distant blazar 1ES 1218+304 (z=0.182) was observed by VERITAS
between December 2006 and March 2007 with a three-telescope
array. This blazar was first detected by MAGIC at TeV
energies \cite{2006ApJ...642L.119A}, albeit with a lower statistical
significance. The more sensitive VERITAS observations allow a
measurement of the spectrum beyond 1 TeV (see Figure \ref{fig:1218})
which opens up the possibility of studying effects of absorption due
to the mid-infrared component of the extragalactic background
light. This can be achieved by examining measured spectra for 1ES
1011-232 and 1ES 0229+200 in the TeV energy regime. Lower limits from
galaxy counts derived from the Hubble Space Telescope deep sky
survey \cite{2000MNRAS.312L...9M} are used in conjunction with various
EBL scenarios to place a lower limit on the spectral hardness of the
three blazars. In particular, the spectrum of 1ES 1218+304 is found to
be harder than $-1.86 \pm 0.37$. A detailed analysis of the intrinsic
spectrum of 1ES 1218+304 will be the subject of a forthcoming
publication. See \cite{Fortin} in these proceedings for further
details.

\section{TeV and X-ray Study of Markarian 421 in 2008}

The nearby blazar Markarian 421 (z=0.03) was observed by VERITAS in an
active state in the first half of 2008. The gamma-ray flux was highly
variable, ranging from 0.3 Crab units to ~10 Crab units \cite{2008ATel.1506....1S}
above 0.5 TeV. 

A total of 93 RXTE PCA pointings and 51 SWIFT XRT observations were
analysed with standard HEAsoft tools. Combined, the SWIFT and XRT data
allow a calculation of the X-ray flux from 0.2-10 keV. It was found
that the data could be fit using both a simple power law and a
log-parabola. The log-parabolic fit yielded a superior $\chi^2$
statistic and fit residuals with no systematic deviations. The
spectral curvature is inconsistent with zero, which suggests the
existence of an intrinsic curvature to the spectrum.

Consistent with previous observations, it is found that Markarian 421
is more variable in the harder X-ray band, and the source hardens
during flaring. Using a discrete correlation function, it is found
that the X-ray and TeV data are correlated with zero lag. This
correlation favours inverse-Compton models for TeV gamma-ray
emission. See \cite{Reyes} in these proceedings for further details.

\begin{figure}
  \includegraphics[height=.3\textheight]{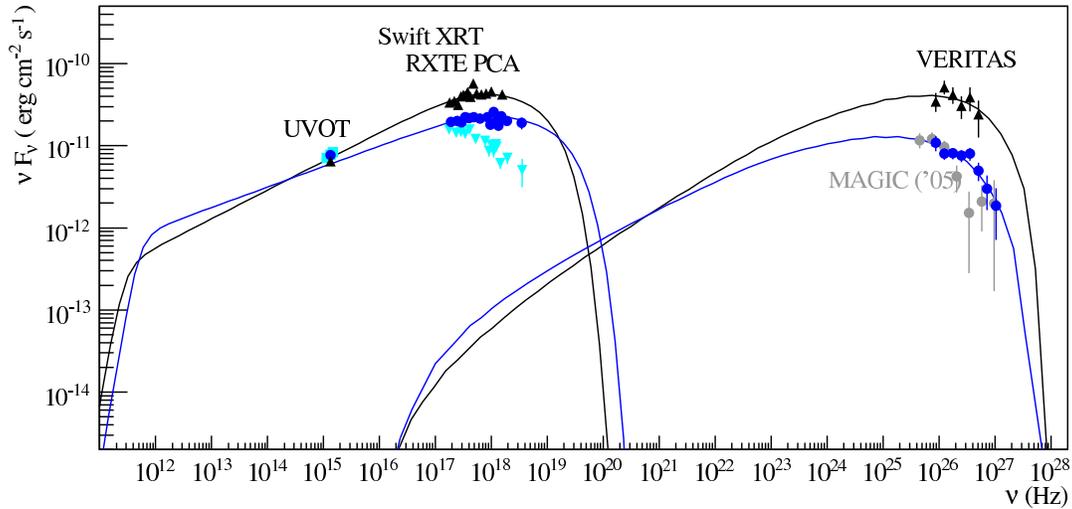} \caption{Spectral
  energy distribution of 1ES 2344+514 comprising SWIFT UVOT and XRT
  observations, RXTE PCA observations, VERITAS and MAGIC
  observations. The data are fit for two periods using a synchrotron
  self-Compton model with reasonable parameters.} \label{fig:2344}
\end{figure}

\section{Broadband Spectral Energy Distribution of 1ES 2344+514}

The nearby blazar 1ES 2344+514 (z=0.044) was observed by VERITAS from
October 2007 to January 2008, resulting in a detection with a
statistical significance of 20.5 sigma with 580 excess events. During
the same period, 1ES 2344+514 was observed by SWIFT XRT for 9
observations and RXTE PCA for 52 observations. SWIFT recorded the
highest known X-ray flux from 1ES 2344+514 on December 8th 2007, with
an X-ray power law index ranging from $2.7 \pm 0.2$ to $1.87 \pm
0.04$. The TeV flux showed variability on a daily timescale, with a
strong flare recorded on 6/7 December 2007 with an integral flux
corresponding to $0.41 \pm 0.05$ of the Crab Nebula flux above 300
GeV. A linear fit with slope $0.7 \pm 0.1$ to the almost simultaneous
X-ray and TeV fluxes suggests a correlation between X-ray and TeV
emission on daily timescales. The broadband spectral energy
distribution of 1ES 2344+514, comprising data from SWIFT UVOT and XRT,
RXTE PCA and VERITAS is shown in Figure \ref{fig:2344}. The data can
be modeled using a one zone synchrotron self-Compton model with
reasonable parameters. See \cite{Grube} in these proceedings for
further details.

\section{Summary and Conclusions}

During the construction phase and first full year of operation, known
and candidate TeV blazars have been the subject of observations by
VERITAS. During this period, the blazars W Comae and 1ES 0806+524 have
been uncovered as sources of TeV gamma-ray emission. W Comae is the
first intermediate-peaked BL Lac object to be discovered. The true
scientific potential of observations in the TeV regime are fully
realised in the context of simultaneous multiwavelength
observations. Data from AAVSO, the SWIFT UVOT and XRT, the RXTE PCA as
well as archival EGRET and radio data have been combined with VERITAS
observations in the TeV regime to produce broadband spectral energy
distributions for W Comae and 1ES 2344+514. These SEDs can be
successfully modeled using leptonic models. Contemporaneous
observations of Markarian 421 with SWIFT, RXTE and VERITAS have allowed
a detailed examination of X-ray/TeV correlations. Finally observations
of 1ES 1218+304 have resulted in a measurement of the spectrum above 1
TeV. This measurement can be used in conjunction with various EBL
scenarios to place a lower limit on the spectral hardness of three
blazars. The first full year of operations of VERITAS have been an
extremely fruitful period. Further results are anticipated from the
exiting observing program.

\begin{theacknowledgments}
This research is supported by grants from the U.S. Department
of Energy, the U.S. National Science Foundation,
and the Smithsonian Institution, by NSERC in
Canada, by PPARC in the UK and by Science Foundation
Ireland.

\end{theacknowledgments}


\begin{thebibliography}{9}




\bibitem[Swordy(2008)]{2008ATel.1415....1S} Swordy, S.\ 2008, The 
Astronomer's Telegram, 1415, 1 

\bibitem[Swordy(2008)]{2008ATel.1422....1S} Swordy, S.\ 2008, The 
Astronomer's Telegram, 1422, 1 



\bibitem[Acciari et al.(2008)]{2008ApJL...648..L73} Acciari, V.~A., et al.\ 
2008, \apjl, 648, L73 


\bibitem[Swordy(2008)]{2008ATel.1565....1S} Swordy, S.\ 2008, The 
Astronomer's Telegram, 1565, 1 


\bibitem[Verrecchia et al.(2008)]{2008ATel.1582....1V} Verrecchia, F., et 
al.\ 2008, The Astronomer's Telegram, 1582, 1 


\bibitem[Beilicke(2009)]{Beilicke} Beilicke, M. \ 2009, These proceedings



\bibitem[Albert et al.(2006)]{2006ApJ...642L.119A} Albert, J., et al.\ 
2006, \apjl, 642, L119 

\bibitem[Madau \& Pozzetti(2000)]{2000MNRAS.312L...9M} Madau, P., \& Pozzetti, L.\ 2000, \mnras, 312, L9 

\bibitem[Fortin(2009)]{Fortin} Fortin, P. \ 2009, These proceedings

\bibitem[Swordy(2008)]{2008ATel.1506....1S} Swordy, S.\ 2008, The 
Astronomer's Telegram, 1506, 1 


\bibitem[Reyes(2009)]{Reyes} Reyes, L. \ 2009, These proceedings



\bibitem[Grube(2009)]{Grube} Grube, J. \ 2009, These proceedings





\end{thebibliography}
\end{document}